\documentclass[12pt]{article}
\begin{document}

\vspace*{.5cm}
\begin{center}
{\bf {\Large {On the parameters of the QCD-motivated potential\\ in the 
relativistic independent quark model }}}
\bigskip

{\large V.V. Khruschev$^{a,c}$, V.I. Savrin$^{b}$ and 
{S.V. Semenov}$^{c,}$}
\footnote{E-mail: semenov@imp.kiae.ru}
\end{center}
\smallskip

$^{a}$Center for Gravitation and Fundamental Metrology, VNIIMS, Moscow, Russia

$^{b}$Institute of Nuclear Physics, Moscow State University, Russia

$^{c}$Russian Research Center ''Kurchatov Institute'',  Kurshatov Square 1,
Moscow,\\
\hspace*{6mm} 123182, Russia

\vspace{.5cm}
\begin{abstract}

In the framework of the relativistic independent quark model the
parameters of the QCD-motivated static potential and 
the quark masses are calculated on the basis of the $1^{--}$ meson 
mass spectra. The value of the confining potential coefficient  
is found to be  $(0.197\pm 0.005)$ GeV$^2$ for  quark-antiquark 
interaction independently on their flavours.  The dependence 
of the quasi-Coulombic potential strength on the interaction distance 
are consistent with the QCD-motivated behaviour. 
The $q\bar q$-separations are evaluated and the $e^+e^-$ decay widths 
are estimated with the help of relativistic modification of the 
Van Royen-Weisskopf formula.

\medskip
\noindent {\it PACS:} 12.39.Pn; 12.39.Ki; 12.40.Yx

\noindent {\it Keywords:} Quark model; Potential; Meson masses; 
Decay widths; Strong coupling constant; String tension; Quark masses; 
Radial excitations
\end{abstract}

\newpage
\medskip

The calculation of spectroscopical hadron characteristics within the data
precision [1] still remains among the unsolved QCD problems. The main 
difficulty consists in consideration of the nonperturbative QCD effects,
which are not  understood yet from first principles but only in terms of
various QCD-inspired models. Thus, at the present time phenomenological
hadron models can be used to a certain extent for the data interpretation.
Among phenomenological hadron models of different kinds 
( e.g.,  Refs. [2-8] ) the relativistic model for 
quasi-independent quarks seems to be one of the most interesting and 
simplest [9]. In our previous papers [10--13] we developed the translation 
invariant version of this model for description of spectroscopical 
characteristics of both the light and heavy vector mesons. We used the Dirac 
equation with a static QCD-motivated potential to describe the quark or 
antiquark motion in the mean field inside the meson. The model does not 
contradict the first principles of QCD and allows one in a simple fashion to 
carry 
out numerical calculations of the meson characteristics, such as their masses, 
the average separations between quarks and antiquarks and the $e^+e^-$ decay 
widths of vector mesons. However, in those papers the effective potential 
entered in the second order model radial equation was chosen for simplicity in
a approximate form. In particular this simplification lead to the 
constancy of the  quasi-Coulombic potential strength 
evaluated on the basis of $1^{--}$ meson mass spectra for different quark 
flavours.  

In the present paper in the framework of the relativistic independent quark 
model  [10--13] we use the exact expression for the static interaction 
effective
potential and evaluate the potential parameters coupled with the quark 
masses on the basis of $1^{--}$ meson mass spectra. Then the obtained values 
are used for calculating the masses of  $1^{--}$ radial excitations.
The $e^+e^-$ decay widths of heavy mesons are estimated with the help 
of relativistic modification of the Van Royen-Weisskopf formula. The 
$q\bar q$-separations for the light and heavy mesons are evaluated as well. 
We find an alternation of the strong coupling constant $\alpha_s$ (which 
depends on the quark flavour and is consistent with the QCD asymptotic 
freedom phenomenon) together with the constancy of the confining potential
coefficient for all quark flavours. The latter result supports the 
hypothesis of the flavor independence  of confinement potential [14--16]. 

According to the main statement of the independent quark model the
hadron is considered as a system composed of the non-interacting with each 
other directly valent constituents (quarks and antiquarks) 
moving in some mean field. One supposes that this field is a colour 
singlet confining 
field, which is produced by the constituents and takes into account the 
effects of creation and annihilation of the sea $q\bar q$ pairs as well. 
To simplify  calculations it is assumed that this  mean field is 
spherically symmetric and its motion in space is determined by motion of its 
center. Furthermore, in order to keep out the ghost-motion states of the
center of mass for the independent constituents, we associate the center of 
mass with the mean field center.  We treat the mean field as 
a quasi-classical object possessing some energy, however one cannot evaluate
the mean field energy without additional assumptions, and this quantity is a 
phenomenological parameter in the  model considered. Each of 
constituents interacting with the spherical mean field gets the state with 
a definite value of its energy. On the phenomenological ground the mass
formula for $J^{PC}$ meson, which is $n^{2S+1}L_J$ state of  
$q\bar q'$-system, can be presented in the following form [10--13]:

\begin{equation}\
M_n(J^{PC})=E(n_r, j)+E(n'_r, j'), 
\end{equation}

\noindent where  $n_r(n'_r)$ and $j(j')$ are the radial quantum number and the
angular moment quantum number of the quark (antiquark). $E(n_r,j)$  is called 
the energy spectral function or the mass term for the quark  and 
in the framework of our model it has got the following phenomenological 
expression: 

\begin{equation}
E(n_r,j)=[\lambda+m^2]^{1/2}+c[1+(-1)^{L+j+1/2}], 
\end{equation}

\noindent  The first term represents the relativistic effective
energy of the constituent moving in the mean field inside the meson, the quark 
mass $m$ being a model parameter and $\lambda$ being found as an eigenvalue of 
the radial  equation of the model. The second term in the formula (2) includes
a part of energy of the mean field and is purely phenomenological one, which
gives small corrections to the spectra of different meson families.

In the equal-time approximation the entire system, the vector $q\bar q$ 
meson, is described by a stationary wave function in the center-of-mass 
frame, which in the  translation-invariant model [10--13] can be 
represented as follows:

\begin{equation}
\Phi_\eta({\bf r}, {\bf r}')=\phi_0\bar\psi({\bf r})e_{\eta}^{\mu}\gamma_{\mu}
U_C\bar\psi({\bf r}'),\label{eq:1}
\end{equation}
where ${\bf r}$ and ${\bf r}'$ are the quark and antiquark coordinates 
in respect to the mean field center, $\phi_0$ is some constant, 
$e_\eta$ is a 4-vector of the meson polarization, $\eta$ describes a 
meson polarization state and takes the values (1, 0, $-1$), $U_C$ is the charge
conjugation operator. Thus, a constituent turns out to be coupled to the other 
via the common mean field which is moving together with their center of mass. 

The wave function $\psi({\bf r})$ for any 
constituent (the quark or antiquark orbital) is a solution of the 
single-particle equation with a static interaction potential. We choose the 
Dirac equation  with the QCD-motivated potential $V(r)$ for description of  
interaction of the quark or antiquark with the external mean field in order 
to determine the orbitals inside the meson. The potential $V(r)$ is 
spherically symmetrical  and consists of the Lorentz scalar and vector 
parts: $V(r)= \beta V_0(r)+ V_1(r)$. Hence the equation for a fermionic 
constituent is 

\begin{equation}
\sqrt{\lambda+m^2}\psi({\bf r})=
\left[(\mbox{\boldmath $\alpha$}{\bf p})+\beta (m+V_0)+V_1\right]\psi({\bf r})
\end{equation}

\noindent  with $V_0(r)=$ $\sigma r/2$ and $V_1(r)=$ $-2\alpha _s/3r$, where
the two model parameters $\sigma $ and $\alpha _s$ are introduced and have
got  meanings of the string tension and the strong coupling constant 
at small distances, correspondingly. 

It is well-known that the solutions of Eq. (4) with the total angular momentum
$j$ and its projection $m$ can be represented as

\begin{equation}
\psi({\bf r})\propto\left(  
\begin{array}{r}
f(r)\Omega _{jl}^m({\bf n}) \\ 
-ig(r)(\mbox{\boldmath $\sigma$}{\bf n})\Omega _{jl}^m({\bf n}) 
\end{array}
\right),
\end{equation}
\noindent  where ${\bf n} ={\bf r}/r$. If $k=-\omega(j+1/2)$ where $\omega$ is 
an eigenvalue of the space-parity operator, the system of the radial Dirac 
equations for the fermion in the mean field with the definite energy sign 
and spin projection reads

\begin{eqnarray}
(\sqrt{\lambda+m^2}-V_0-V_1-m)f =-\frac{(k+1)}{r}g-g',  \nonumber \\
(\sqrt{\lambda+m^2}+V_0-V_1+m)g =-\frac{(k-1)}{r}f+f'.  \label{6}
\end{eqnarray}

Using Eqs. (6) one can derive the second order equation
for the "large" component $f(r)$, and then making a substitution

\begin{equation}
\varphi(r)=rf(r)\left[ V_0(r)-V_1(r)+m+\sqrt{\lambda+m^2}\right]^{-1/2} 
\end{equation}
one comes on to the model radial equation for $\varphi(r)$ in the following 
form: 

$$
\varphi^{^{\prime \prime}}+\lambda\varphi=
\left [
(m+V_0)^2-(\sqrt{\lambda+m^2}-V_1)^2+
\frac{k(k-1)}{r^2}+\frac{3(V_0^{^{\prime }}-V_1^{^{\prime }})^2}
{4(\sqrt{\lambda+m^2}-V_1+V_0+m)^2}
\right.
$$
\begin{equation}
\left.
-\frac{k(V_0^{^\prime}-V_1^{^{\prime }})}{r(\sqrt{\lambda+m^2}-V_1+
V_0+m)}-\frac{(V_0^{^{\prime \prime }}-V_1^{^{\prime \prime }})}
{2(\sqrt{\lambda+m^2}-V_1+V_0+m)}
\right ]
\varphi. 
\end{equation}
\smallskip

Further on we restrict ourselves evaluating only characteristics of the radial 
excitations of the S-wave $1^{--}$ mesons because they are described by the 
simplest version of model and supported by the most extensive set of accurate 
data [1], especially for heavy mesons. So, $k=1$ and the contribution of the  
term containing $c$ in Eq. (2) is equal to zero. $\lambda$ entering Eq. (2) can
be calculated with the help of the S-wave radial model equation: 

$$
\varphi^{^{\prime \prime}}+\lambda\varphi =
\left \{
-\frac{4\alpha_s\sqrt{\lambda+m^2}}{3r}-\left(\frac{2\alpha _s}{3r}\right)^2+
m\sigma r 
\right.
$$
\begin{equation}
\left.
+\left(\frac{\sigma r}{2}\right)^2-\left[\frac{\sigma r}{2}\left(m+
\sqrt{\lambda+m^2}\right)+\left(\frac{\sigma r}{4}\right)^2+
\frac{5\alpha _s\sigma}{6}
-\frac{\alpha _s^2}{3r^2}\right]
\right / 
\end{equation}
$$
\left.
\left[\frac{2\alpha _s}{3}+r\left(m+\sqrt{\lambda+m^2}\right)+
\frac{\sigma r^2}{2}\right]^2
\right\}
\varphi. 
$$
\smallskip

The right-hand side of  Eq. (9) has a singularity at the origin, and when
$r\to 0$ it behaves as $3/4r^2-4\alpha _s^2/9r^2$. Therefore one should 
keep $\alpha_s<3/2$ in order to prevent a fall at the origin. At the 
infinity $r\rightarrow \infty $ the effective potential behaves as the 
oscillatory one and tends to $\sigma^2r^2/4$.

In the framework of our model the quantities  $m$, $\sigma $ and $\alpha _s$ 
are phenomenological parameters and to be determined as a result of 
calculations of meson mass spectra, as well as the $e^+ e^-$ decay widths, 
and a comparison of them with experimental data.

The model equation (9) can be solved only by  numerical methods. 
For calculating its eigenvalues we used the computer code algorithm, which
is based on the Numerov three point recurrent relation for the equation
$y^{\prime \prime }=F(r,E)y$ [17]:

\begin{eqnarray}
y_{i-1} =\left[ y_i\left( 2+\frac 56F(r_i,E)h^2\right) -y_{i+1}\left( 1-
\frac{F(r_{i+1},E)h^2}{12}\right) \right]  
\times\left[ 1-\frac{F(r_{i-1},E)h^2}{12}\right] ^{-1}.  \label{eq:7}
\end{eqnarray}
For bound states  $y(r)$ tends to zero when  $r > r_{cl}$. Here $r_{cl}$ is a 
classical radius of the bound state, which is determined from the equation 
$U(r_{cl})=E$ and $E$ is the initial energy of the considered level. Then the 
$E$ value  can be determined from the condition: $y(0)=0$. However,  $F(r,E)$ 
has a singularity at the origin. So, when we calculate $y(0)$ with the help of 
formula (10) we use a regularization procedure for the singular potential, 
which is analogous to the procedure worked out, for instance, in Ref. [18].

In order to estimate the precision of the calculation algorithm the well-known 
test with diminished spacing $h$ is used. The results obtained allow us to 
disregard machine calculation errors as compared with systematic errors of the 
model. The accuracy criterion for fitted parameters was the maximum value of 
the acceptable errors for  evaluated hadron mass values as compared with 
typical experimental errors which we choose to be in the range of $30\div 40$ 
MeV. Thus, the parameter errors written below must be considered as our 
estimation of systematic errors of the model.

When calculating the values of the model parameters we do not suppose 
{\it a priori} the validity of the hypothesis of flavor independence for the
confining potential. First of all the  model parameters $m$, $\alpha _s$ and 
$\sigma $ for the $b\bar b$ and $c\bar c$ radial excitations of $1^{--}$ states 
were found. The fit was carried out for each meson family independently. The 
existing experimental data allowed to find these parameters as well as to prove
the validity of the model. In this manner it was found that the value of 
$\sigma $ is the same for the $b\bar b$ and $c\bar c$ states within the 
systematic errors of the model and equal to $(0.197\pm 0.005)$ GeV$^2$. Taking 
into account that there are no well established data for the higher radial 
excitations of $1^{--}$ light mesons, the  obtained $\sigma $ value  was used 
for calculating  $m$ and $\alpha_s$ for light quarks. Note that when doing 
the calculations we neglect the isotopic mass splitting of the $u$- and 
$d$-quarks because  it is beyond of the model accuracy.

All results for the well known experimental data on 
the $1^{--}$ mesons, which are composed of the quark and antiquark 
with $u$-, $d$-, $s$-, $c$- or $b$-flavours, are consistent 
within the model accuracy with the following 
values for the coupling $\alpha _s$ (quasi-Coulombic potential strength) and 
the quark masses:

\begin{center}
\begin{tabular}{ll}
${\bar m}_{u,d}=( 0.01 \pm 0.008)$ GeV, & $\alpha_s^{u,d}=0.70\pm 0.15,$ \\ 
$m_s=(0.18\pm 0.03)$ GeV, & $\alpha_s^s=0.50 \pm 0.10,$ \\ 
$m_c=(1.35\pm 0.05)$ GeV, & $\alpha_s^c=0.36 \pm 0.03,$ \\ 
$m_b=(4.70\pm 0.10)$ GeV, & $\alpha_s^b=0.24 \pm 0.02.$
\end{tabular}
\end{center}

Now let us  estimate the  $ e^{+}e^{-}$  decay widths of the vector
mesons  with the help of solutions of the Dirac equations (6) which were 
obtained numerically. It is well known that in the nonrelativistic potential 
approach the vector meson $ e^{+}e^{-}$   width is described by the  
Van Royen-Weisskopf formula with  first order QCD correction:

\begin{equation}
\Gamma_{_V}=\frac{16\pi\alpha^2e_q^2}{M_{_V}^2}|\Psi(0)|^2
\left(1-\frac{16\alpha_s}{3\pi}\right), \label{eq:16}
\end{equation}

\noindent where $|\Psi(0)|^2$ is  the square of the meson wave 
function, which describes the probability density of meson at the origin. 
We shall keep a structure of the formula (\ref{eq:16}) 
in the relativistic independent quark model. 

Using Eqs. (3) and (5) the wave function of the meson consisting of independent
quark and antiquark in the $S$-state can be written as

\begin{equation}
\Phi_\eta({\bf r},
{\bf r}')=\frac{\phi_0}{4\pi}\left[-f(r)f(r')\stackrel{\ast}
{w}(\mbox{\boldmath $\sigma$}{\bf e}_\eta)w'+g(r)g(r')\stackrel{\ast}{w}
(\mbox{\boldmath $\sigma$}{\bf n})
(\mbox{\boldmath $\sigma$}{\bf e}_\eta)(\mbox{\boldmath $\sigma$}{\bf
n}')w'\right],\label{eq:13}
\end{equation}
where we have  used a relationship $\sigma_2\stackrel{\ast}
{\mbox{\boldmath $\sigma$}}=-\mbox{\boldmath $\sigma$}\sigma_2$ and denoted 
$w'=-\sigma_2\stackrel{\ast}{w}$ (the asterisk denotes complex conjugation).
So the probability density of independent quark-antiquark system can be easily 
calculated: 
$$
|\Phi_\eta({\bf r}, {\bf r}')|^2=\frac{2\phi_0^2}{16\pi^2}\big\{f^2(r)
f^2(r')+g^2(r)g^2(r')+2f(r)g(r)f(r')g(r')({\bf n}{\bf n}')-
$$
\begin{equation}
-2f(r)g(r)f(r')g(r')[({\bf e}_\eta{\bf n})(\stackrel{\ast}{\bf e}_\eta{\bf n}')
+(\stackrel{\ast}{\bf e}_\eta{\bf n})({\bf e}_\eta{\bf n}')]\big\}.
\label{eq:14}
\end{equation}
And the averaged over initial spin projections probability density is
\begin{equation} 
|\overline{\Phi_\eta({\bf r}, {\bf r}')}|^2 =\frac{2\phi_0^2}{16\pi^2}
\big[f^2(r)f^2(r')+g^2(r)g^2(r')+2/3f(r)g(r)f(r')g(r')({\bf n}{\bf n}')\big].
\label{eq:15}
\end{equation}

Now taking into account that $\bf r$ and ${\bf r}'$ are the coordinates of the 
quark and the antiquark in respect to their center of mass, which coincides 
with the center of the mean field, we can define according to Eq. (\ref{eq:15}) 
the probability density of meson in the following way:

\begin{equation}
\int\!\! d^3{\bf r}'\;|\overline{\Phi_\eta({\bf r}, {\bf r}')}|^2 =
\frac{1}{2\pi}\left[\kappa_1 f^2(r)+ \kappa_2 g^2(r)\right],
\label{eq:17}
\end{equation}
where we denote
\begin{equation}
\int\!\!drr^2\; f^2(r)=\kappa_1, \quad \int\!\! drr^2\; g^2(r)=\kappa_2.
\label{eq:18}
\end{equation}

Hence the full probability is

\begin{equation}
\int\!\! d^3{\bf r}\!\!\int\!\!
d^3{\bf r}'\;|\overline{\Phi_\eta(\bf{r}, \bf{r}')}|^2 =
2\left(\kappa_1^2 + \kappa_2^2\right).\label{eq:19}
\end{equation}

\noindent and therefore the normalized propability density which is analogous
to the  $|\Psi(r)|^2$ is equal to $\left(\kappa_1 f^2(r)+ \kappa_2 g^2(r)
\right)/4\pi \left(\kappa_1^2 + \kappa_2^2\right)$. This quantity at $r=0$ must
enter the formula (\ref{eq:16}) instead of  the  $|\Psi(0)|^2$. However,
in relativistic  theory one cannot localized a quark within the volume
with the radius less than 
 the Compton wave length of the quark $\lambda_q\cong m_q^{-1}$.  According to 
this reasoning we use for the estimation of decay width the following formula:

\begin{equation}
\Gamma_{V_n}=\frac{4\alpha^2e_q^2}{M_{V_n}^2}\left[\frac{\kappa_1
f^2_n(\lambda_q)+\kappa_2 g^2_n(\lambda_q)}{\kappa_1^2 +
\kappa_2^2}\right]
\left(1-\frac{16\alpha_s}{3\pi}\right).
\label{eq:20}
\end{equation}

Taking into account the expression for the normalized probability density,
the mean squared radius of the meson is 

\begin{equation}
R^2_n=\int\!\!drr^4\; \left[\frac{\kappa_1
f^2_n(r)+\kappa_2 g^2_n(r)}{\kappa_1^2 +
\kappa_2^2}\right].
\label{eq:21}
\end{equation}

The results obtained \marginpar{\fbox{Table 1 }}for the meson masses 
$M_{n}^{th}$ and the average radii $R_n^{th}$ are displayed in  Table 1. 
We estimate  the $e^{+}e^{-}$ decay widths  $\Gamma_{n}^{th}$
with the help of formula (\ref{eq:20})
only for the $c\bar c$- and $b\bar b$-mesons because there are no
satisfactory data for the radial excitations of light mesons. 
The mass values $M_n^{th}$ and  $M_n^{exp}$ coincide with each other within
the model accuracy. Taking into account the approximate nature of the formula
(\ref {eq:20}) there is a conformity between the calculated decay 
ratios and the experimental values [1]. However, some futher efforts are needed 
in order to improve  the precision for decay widths calculations
in the framework of model considered. Note that there is a  good 
agreement between the $R_n^{th}$ and the values calculated in the framework 
of the lattice QCD in the valence quark approximation [19]. 
\medskip 

In this paper we considered the relativistic hadron model with independent
quarks and evaluated the parameters of the $q\bar q$ model potential, as well
as the masses, the average radii of the $1^{--}$ mesons, and estimated the 
$e^+e^-$ decay widths. We find that the value of $\sigma=(0.197\pm 0.005)
$ GeV$^2$ is the admissible string tension value for $u$-, $d$-, $s$-, $c$- and
$b$-quark flavours within the systematical errors of the model. It is an 
important result of our paper which confirms the flavor independence of 
confining potential on the $3\cdot 10^{-2}$ level of accuracy. Thus one may 
interpret  $\sigma$ as the universal parameter of quark-antiquark interaction 
at large distances and consider the characteristic confinement scales such as 
the confinement length $l_c$ and the confinement mass $m_c$, which satisfy the 
following relations: $m_cl_c=1$, $\sigma= m^2_c$.

In addition, the correspondence to the QCD asymptotic freedom phenomenon, or  
the decrease of $\alpha _s$ when the interaction range diminishes, is confirmed
both for the light and heavy quarks in  the framework of this model. Note that 
within  the large systematical errors of the model 
it is difficult to prove the logarithmic dependence of $\alpha _s$ on $Q^2$,
moreover, there are  considerable nonperturbative contributions for the light 
quarks (see, e.g. Refs. [20, 21]). Nevertheless, there is no contradiction 
between the $\alpha _s(Q^2)$ behaviour in QCD and the diminution of the 
parameter $\alpha_s$ in this model when coming from the heavy to the 
light quarks.

An advantage of the presented model is its conceptual clarity and
technical simpli\-ci\-ty, which allows one to calculate in a simple fashion the 
meson characteristics such as  masses,  $q \bar q$-separations  
and  $ e^{-}e^{+}$  decay widths with an admissible among phenomenological 
models accuracy. For instance, the values of the parameters $m_b$ and $m_c$
evaluated in the framework of this model are similar to the values of
the pole masses $M_b$ and $M_c$ obtained in Ref. [7].
The values of 
$\alpha_s(m_i^2)$ ($i=b,\; c$)  are close to the  $\alpha_s(M_i^2)$ 
in the $\overline {MS}$ scheme, whereas 
the average $q\bar q$-separations for  heavy vector mesons are in agreement 
with those found on the basis of the lattice QCD [19].
Moreover, in this paper the calculations of $q\bar q$-separations for light 
vector mesons 
are presented, while the same calculations are not possible now in the
frame of the lattice QCD. In order to increase the model accuracy it is 
reasonable to take into account the relativistic corrections due to contact 
particle-particle interaction, like those which have been treated in Ref. [2].
The account for such kind of corrections lies beyond the scope of the 
independent
quark model and will  be of our further task in order to improve the model
accuracy and to find its limitations. Besides we intend to apply
this model to the more complicated systems, such as the baryons and
the exotic multiquark mesons.

\smallskip\ \noindent {\bf Acknowledgments}. The authors express their
gratitude to R.N.~Faustov and A.L.~Ka\-ta\-ev for useful discussions.

\newpage

{\noindent \small Table 1. \quad{} 
\parbox[t]{13.9cm}{Evaluated masses and $e^+e^-$ decay widths
 of the $1^{--}$ mesons in comparison with the  data from Ref. [1], and the 
 average $q\bar q$-separations in comparison with the results of Ref. [19].}}

\begin{center}
\begin{tabular}{|c|c|c|c|c|c|c|}
\hline
Meson & $M_n^{exp} [MeV]$ & $M_n^{th} [MeV]$ & $R^{th}_n[fm]$
&$R_n^{[19]}[fm]$ & $\Gamma_n^{th}[keV]$ & $\Gamma_n^{exp}[keV]$\\ \hline
$\rho $ & 769.9$\pm $0.8 & 740 & 0.65 &-&-&6.77$\pm$ 0.32\\ 
$\rho ^{\prime }$ & 1465$\pm$25 & 1455 & 1.12 &-&-&-\\ 
$\rho ^{\prime\prime }$ & 1700$\pm$20 & 1730 & 1.32 &-&-&-\\ 
$\phi $ & 1019.413$\pm $0.008 & 1010 & 0.58 &-&-&4.43$\pm$0.05\\ 
$\phi ^{\prime }$ & 1680$\pm $50 & 1650 & 0.96 &- &-&-\\ 
$\phi ^{\prime\prime }$ & - & 2050 & 1.34 &- &-&-\\ 
$J/\psi $ & 3096.88$\pm $0.04 & 3060 & 0.41 & 0.43
&7.32&$5.26\pm0.37$\\ 
$\psi ^{\prime }$ & 3686.00$\pm $0.09 & 3650 & 0.79 & 0.85
&2.03&$2.14\pm0.21$\\ 
$\psi ^{\prime \prime }$ & 4040$\pm $10 & 4070 & 1.06 & 1.18
&0.97&$0.75\pm0.15$\\ 
$\psi ^{\prime \prime \prime }$ & 4415$\pm $6 & 4390 & 1.30 & 1.47
&0.58&$0.47\pm0.10$\\ 
$\Upsilon $ & 9460.37$\pm $0.21 & 9470 & 0.26 & 0.24&1.30&$1.32\pm0.05$\\ 
$\Upsilon ^{\prime }$ & 10023.3$\pm $0.3 & 9990 & 0.54 & 0.51
&0.37&$0.520\pm0.032$\\ 
$\Upsilon ^{\prime \prime }$ & 10355.3$\pm $0.5 & 10325 & 0.75 & 0.73 
&0.21&-\\ 
$\Upsilon ^{\prime \prime \prime }$ & 10580$\pm $3.5 & 10550 & 0.94
& 0.93&0.16&$0.248\pm0.031$\\ 
$\Upsilon ^{5S}$ & 10865$\pm $8 & 10830 & 1.10 & - &0.12&$0.31\pm0.07$\\ 
$\Upsilon ^{6S}$ & 11019$\pm $8 & 10985 & 1.25
& -&0.08&$0.13\pm0.03$\\ \hline
\end{tabular}
\end{center}

\end{document}